\documentclass[aps,prl,reprint,preprintnumbers]{revtex4-1}
\pdfoutput=1 
\usepackage{blindtext}
\usepackage{array}
\usepackage{amsmath}
\usepackage{amsfonts}
\usepackage{amssymb}
\usepackage{amsthm}

\usepackage[english]{babel}
\usepackage[T1]{fontenc}
\usepackage{lmodern}

\usepackage{comment}
\usepackage{tikz}
\usetikzlibrary{decorations.markings}
\usepackage{bbm}

\usepackage{subcaption}  
\usepackage{caption}
\usepackage{graphicx,subcaption}

\usepackage{hyperref}
\hypersetup{
   colorlinks=true,
   urlcolor=blue,    
   linkcolor=black,  
   citecolor=blue,   
}

\begin{document}
\preprint{HU-EP-25/01-RTG, MITP-25-001, MPP-2025-2, USTC-ICTS/PCFT-25-01}
\title{Complete function space for planar two-loop six-particle scattering amplitudes}
\author{Johannes Henn$^{a}$}
\email{Corresponding author. E-Mail: henn@mpp.mpg.de}
\author{Antonela Matija\v{s}i\'{c}$^{a,b}$}
\author{Julian Miczajka$^{a}$}
\author{Tiziano Peraro$^{c}$}
\author{Yingxuan Xu$^{d}$}
\author{Yang Zhang$^{e,f}$}
%
\affiliation{$^a$ Max-Planck-Institut f\"{u}r Physik, Werner-Heisenberg-Institut, Boltzmannstr. 8,
85748 Garching, Germany \\
$^b$ PRISMA Cluster of Excellence, Institut für Physik, Staudinger Weg 7, Johannes Gutenberg-Universität Mainz,  55099 Mainz, Germany \\
$^c$ Dipartimento di Fisica e Astronomia, Universitá di Bologna e INFN, Sezione di Bologna, via Irnerio 46, 40126 Bologna, Italy \\
$^d$ Institut für Physik, Humboldt-Universität zu Berlin, 10099 Berlin, Germany \\
$^e$ Interdisciplinary Center for Theoretical Study, University of Science and Technology of China, Hefei, Anhui 230026, China\\
$f$ Peng Huanwu Center for Fundamental Theory, Hefei, Anhui 230026, China
}
\begin{abstract}
We derive the full system of canonical differential equations for all planar two-loop massless six-particle master integrals, and determine analytically the boundary conditions. This fully specifies the solutions, which may be written as Chen iterated integrals. We argue that 
this is sufficient information for evaluating any scattering amplitude in four dimensions up to the finite part. 
We support this claim by reducing, for the most complicated integral topologies, integrals with typical Yang-Mills numerators.
We use the analytic solutions to the differential equations, together with dihedral symmetry, to provide the full solution space relevant for two-loop six-particle computations. 
This includes the relevant function alphabet, as well as the independent set of iterated integrals up to weight four. 
We also provide the answer for all master integrals in terms of iterated integrals that can be readily evaluated numerically. As a proof of concept, we provide a numerical implementation that evaluates the integrals in part of the Euclidean region, and validate this against numerical evaluation of the Feynman integrals.
Our result removes the bottleneck of Feynman integral evaluation, paving the way to future analytic evaluations of six-particle scattering amplitudes.
\end{abstract}

\maketitle

\section{Introduction}

About ten years ago, obtaining phenomenological predictions for two-to-three collider processes beyond the next-to-leading order was widely considered to be impossible. However, this changed with breakthroughs in our ability to evaluate the complicated two-loop Feynman integrals and to think about the results in terms of Chen iterated integrals \cite{Goncharov:2010jf}. This involved many steps: deriving suitable differential equations for the master integrals, solving them in the relevant physical regions, assembling full scattering amplitudes leveraging finite field ideas, etc. This is nicely reviewed in the recent thesis \cite{zoia2022modern}. Indeed, the last years have seen tremendous progress: analytical calculations for two-loop five-point massless integrals have been achieved \cite{Gehrmann:2015bfy,Gehrmann:2018yef,Abreu:2018aqd,Chicherin:2018old,Chicherin:2020oor}, enabling important phenomenological applications, as demonstrated in \cite{Abreu:2021oya,Chawdhry:2021mkw,Agarwal:2021vdh,Czakon:2021mjy}. Furthermore, recent efforts have extended these findings to processes involving one off-shell leg, with notable studies such as \cite{Canko:2020ylt, Chicherin:2021dyp,Kardos:2022tpo,Abreu:2023rco}
for the master integrals, and \cite{Badger:2021nhg} for scattering amplitudes. In the present paper, we take an important step into the next frontier: we provide the full information on the master integrals and their analytic solutions relevant for planar, massless two-loop six-particle scattering processes. We also provide a proof-of-principle numerical evaluation.

Beyond its phenomenological implications, this analytic information also is of great theoretical interest. 
There has been a resurgence of interest and new mathematical ideas in the Landau method for predicting the singular locus of Feynman integrals \cite{Dlapa:2023cvx,Fevola:2023kaw,Fevola:2023fzn}, which is a way of constraining the function space. Moreover, starting from results in maximally supersymmetric Yang-Mills theory (sYM), several studies have looked into possible cluster algebra interpretations of function spaces \cite{Golden:2013xva,Chicherin:2020umh,Henke:2021ity}. Our six-particle results for the function alphabet provide a highly non-trivial data point and benchmark for these ideas. Moreover, work in sYM has also revived interest in the analytic structure of amplitudes, such as the Steinmann relations \cite{Caron-Huot:2016owq}, adjacency properties, crossing relations \cite{Caron-Huot:2023ikn}, questions of factorization in certain channels \cite{Henn:2024qjq} and intriguing antipodal relations \cite{Dixon:2021tdw}. By providing the complete function space, our work opens the possibility of investigating similar questions for generic Yang-Mills theories.

\begin{figure}[!t]
    \centering
\includegraphics[width=\linewidth]{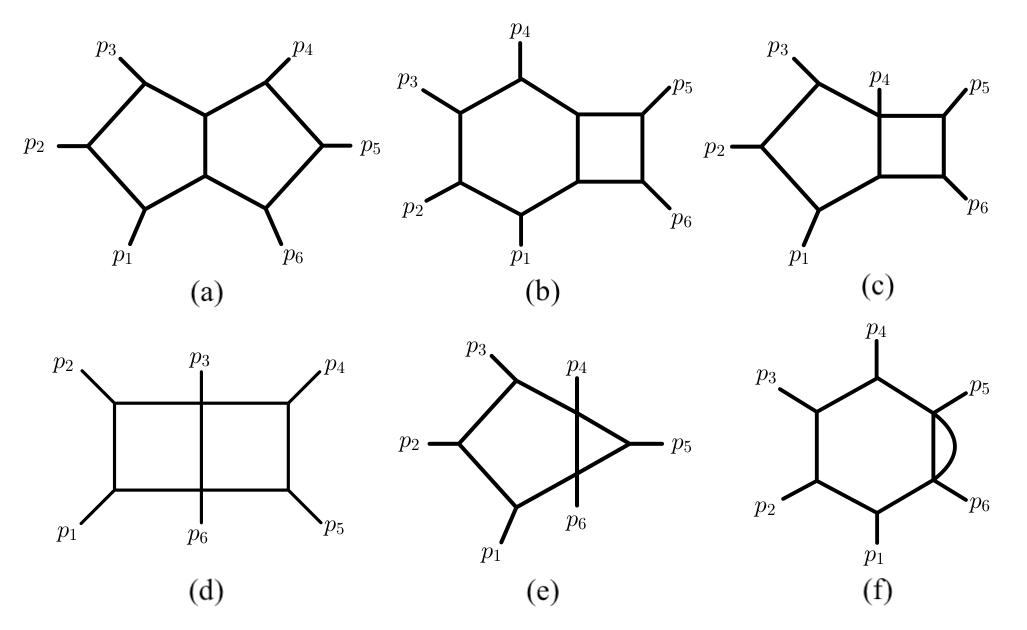}
    \caption{Genuine six--point two--loop 
    planar Feynman diagrams.}
    \label{fig:families}
\end{figure}
We use the fact that all integrals relevant to Yang-Mills amplitudes can be reduced to scalar integrals.
These scalar integrals are further reduced, using integration-by-parts (IBP) identities, to a 
set of basis integrals, which we call master integrals (MI)~\cite{TKACHOV198165, Chetyrkin:1981qh}. The computation of these integrals is performed using  
differential equations~\cite{KOTIKOV1991123,Bern:1993kr,Gehrmann:1999as}.
The latter take a greatly simplified, canonical form \cite{Henn:2013pwa}, if one chooses a basis of uniform transcendental (UT) weight integrals \cite{Arkani-Hamed:2010pyv}.

We use this approach for most of the relevant integrals. However, for the most complicated integrals, of the type shown in Fig.~\ref{fig:families}(a), we proceed differently. Specifically, we show that the relevant integrals for computing two-loop amplitudes up to and including the finite part can be related to integrals known in the literature.

\section{Notation and kinematics}

We consider six-particle scattering in $D=4- 2\epsilon$ dimensional Minkowski space ${\mathbb M}^{D}$. The six external momenta $p_{i} \in {\mathbb M}^{D}$, $i=1\,,\ldots\,,6$, are on-shell, i.e. $p_{i}^2=0$ and satisfy momentum conservation, $\sum_{i=1}^{6} p_i = 0$.
For planar integrals, it is useful to introduce dual coordinates.
The momenta form a closed polygon, whose vertices are dual coordinates $z_i$. In other words,
\begin{align}
z_{i+1} - z_{i} = p_{i} \,,
\end{align}
with $z_7 \equiv z_1$. Moreover, we introduce dual coordinates $y_1, y_2$ corresponding to the loop momenta.
Any scalar planar two-loop hexagon integral can then be written as
\begin{align}
\begin{split}
    I_{\vec{a}} &= e^{2 \epsilon \gamma_E} \int \frac{d^{D}y_1}{i \pi^{D/2}}
    \frac{d^{D}y_2}{i \pi^{D/2}} 
    \frac{1}{[-(y_1 - y_2)^2]^{a_{13}}} \times \\
    & \quad\quad\quad \frac{1}{\prod_{j=1}^{6}  [-(z_j-y_1)^2]^{a_j}  [-(z_j-y_2)^2]^{a_{j+6}} } \,.
    \end{split} \label{eq:generalintegral}
\end{align}
Here $\gamma_E$ is the Euler-Mascheroni constant, and $\vec{a}$ is a vector of $13$ elements that correspond to the powers of the propagator factors. The latter can take positive or negative values.
To translate back to usual momentum space, one may fix e.g. $z_1 =0$, and use $z_{1}-y_{1} = l_1$, $z_{1}-y_{2} = l_2$ for the loop momenta.

For kinematics in $D$ dimensions, the integrals $ I_{\vec{a}}$ depend on nine independent scalar Poincaré invariants, which we chose as
\begin{align}\label{scalarPoincareinvariants}
\{ z_{13}^2, z_{24}^2, z_{35}^2, z_{46}^2, z_{15}^2, z_{26}^2, z_{14}^2, z_{25}^2, z_{36}^2 \} \,,
\end{align}
where we abbreviated $z_{ij} \equiv z_i - z_j$.
For example, $z_{13}= (p_1 + p_2)^2$ and $z_{14}^2 = (p_1 + p_2 +p_3)^2$.

In the following, we will take the external kinematics to lie in a four-dimensional subspace of $D$. 
In this case, the five momenta $p_1, \ldots p_5$ become linearly dependent, which translates into a constraint on the invariants (\ref{scalarPoincareinvariants}).
It is advantageous to use momentum twistor variables \cite{Hodges:2009hk}, which are unconstrained four-dimensional variables (i.e., they also solve the on-shell and momentum conservation conditions). We choose the particular momentum twistors parametrization of section 2 of \cite{Henn:2024ngj}, where the eight independent variables are denoted by $\vec{x}$.

There is also one pseudoscalar Poincaré invariant, which we introduce in order to fully specify the scattering kinematics. We choose it as
\begin{equation}
    \epsilon_{1234}\equiv 4 \sqrt{-1} \ \varepsilon_{\mu_1 \mu_2 \mu_3
    \mu_4} p_1^{\mu_1} p_2^{\mu_2}p_3^{\mu_3}p_4^{\mu_4} \,,
\end{equation}
where $\varepsilon$ is a totally antisymmetric tensor.
Moreover, we define the Gram determinant 
\begin{gather}
    G\left(
\begin{array}{ccc}
u_1 & \dots & {u_n} \\
v_1 & \dots & {v_n} \\
\end{array}
\right)=\mathrm{det}(2 u_i\cdot v_j)\,,
\end{gather}
where the right--hand side is the determinant of the $n\times n$ matrix with entries $2(u_i\cdot v_j),1\leq i,j \leq n$. We also introduce the abbreviation 
\begin{gather}
    G(i_1,\dots,i_k)\equiv     G\left(
\begin{array}{ccc}
p_{i_1} & \dots & p_{i_k} \\
p_{i_1} & \dots & p_{i_k} \\
\end{array}
\right).
\end{gather}

\section{Integral family and the canonical differential equation}

The full set of planar two-loop integral families with genuine six-particle kinematics is displayed in Fig.~\ref{fig:families}. All planar two-loop six-point integrals belong to the double pentagon (DP) in  Fig.~\ref{fig:families}(a) or the hexagon box (HB) in Fig. \ref{fig:families}(b), or to permutations thereof. 

In our previous work~\cite{Henn:2024ngj}, the integral families shown in Figs.~\ref{fig:families}(d),~\ref{fig:families}(e) and~\ref{fig:families}(f) were evaluated, and in ref.~\cite{Matijasic:2024too}, the canonical differential equations for integral families shown in Figs.~\ref{fig:families}(b) and~\ref{fig:families}(c) were determined.

In the present work, we determine the uniform transcendental (UT) weight integral bases for the full DP and HB sectors consisting of 267 and 202 master integrals, respectively, and present the analytic solution up to weight four.
We chose a basis of master integrals $\vec{I}$ such that they satisfy canonical differential equations \cite{Henn:2013pwa}, i.e. 
\begin{equation} 
    \text{d}\vec{I}\,^\text{fam}(\epsilon, \vec{x}) = \epsilon \, \text{d}\tilde A^\text{fam}(\vec{x}) \cdot \vec{I}\,^\text{fam}(\epsilon, \vec{x}), 
    \label{eq:CDE}
\end{equation}  
where $\tilde A^\text{fam}$ is a $267\times 267$ matrix for $\text{fam}=\text{DP}$, a $202\times 202$ matrix for $\text{fam}=\text{HB}$. 
We derived the differential equations using momentum twistor variables in the form of~\cite{Badger:2013gxa},
and
verified the $\epsilon$-factorized form of eq. (\ref{eq:CDE}) using finite field methods.

In order to achieve the canonical form (\ref{eq:CDE}), we chose the top sector for the DP family as in reference \cite{Henn:2021cyv}. 
For the HB and pentagon-box (PB) families, we used the Baikov analysis as in \cite{Dlapa:2021qsl} in order to choose a simpler basis compared to the one used in \cite{Henn:2021cyv}.

In principle one could 
perform a full analytic reduction
to determine the matrix $\tilde{A}$ in eq. (\ref{eq:CDE}). However, 
if one knows the form of the entries of this matrix in advance, one may
use finite field methods~\cite{vonManteuffel:2014ixa,Peraro:2016wsq} to reconstruct them, as in our previous work \cite{Henn:2024ngj}. 
 This significantly lowers the computational cost and speeds up the computation.
So let us assume that the connection matrix of the canonical differential equations \eqref{eq:CDE} 
is expanded in terms of logarithms, i.e.  
\begin{equation}\label{Atildeexplicit}
    \tilde{A}(\vec{x})=\sum_j c_j  \log(\alpha_j(\vec{x})),
\end{equation}
where $c_{j} \in {\mathbb{Q}}$, and the so-called alphabet letters $\alpha_j(\vec{x})$ are algebraic functions of the kinematic variables. 
Our ansatz for the alphabet involves
known letters from the literature for five-point one-mass integrals~\cite{Abreu:2020jxa}, supplemented it with additional algebraic letters that we constructed using the method outlined in~\cite{Heller:2019gkq,zoia2022modern,Matijasic:2024too}, using the computer algebra implementation~\cite{Matijasic:2024gkz}.

In this way, we were able to construct the full matrix $\tilde{A}^\text{fam}$ for the DP and HB families, except for entries associated to one six-dimensional double-pentagon integral. 
The UT integral basis and the canonical differential equation matrix are provided as auxiliary files.
The missing matrix elements, which may or may not be of the form of eq. (\ref{Atildeexplicit}), could in principle be reconstructed using a full reduction. However,
as we discuss presently, the information we have already computed is all that is needed for two-loop amplitude computations up to and including the finite part.

\section{Relevant functions for 4D Yang-Mills scattering amplitudes}

Let us explain why no new information is needed for the most complicated double pentagon sector shown in Fig.~\ref{fig:families}(a).
There are five master integrals \cite{Henn:2021cyv}, i.e. in principle up to five new functions to be computed. However, our goal is to compute two-loop amplitudes up to the finite part only. This allows us to truncate the expansion in $\epsilon$ of the basis integrals. The argument uses the following ingredients:
\begin{enumerate}
\item Two-loop amplitudes in four dimensions involve  terms up to transcendental weight four only \cite{Hannesdottir:2021kpd}.
\item The canonical form 
of the differential equations makes the transcendental weight of the basis integrals manifest. 

\end{enumerate}
We argue that it is sufficient to expand the basis integrals up to weight four. 
Note that for a random basis, this is in general not enough, as the reduction coefficients may in general involve poles in $\epsilon$, which in turn would require us to expand the basis integrals to higher orders (cf. e.g. the discussion in \cite{Gehrmann:2000xj}).
However, we argue that this does not occur for uniform weight integrals (that have the same weight as the amplitude): indeed, if a UT integral with a nonzero $\epsilon^k$ order  meets a pole $\mathcal{O}(1/\epsilon^k)$, this would yield a contribution of weight higher than four to the finite part of the amplitude, in contradiction with point 1 above. We support this statement by computing reductions to master integrals for a complete set of integrals up to rank six that are expected to appear in Yang-Mills theories and reconstructing their dependence on the dimensional regulator $\epsilon$ on the maximal cut.  We have also extended this analysis without cut for a selection of typical integrals contributing to Yang-Mills amplitudes.  We found that all coefficients of these reductions are finite in the limit $\epsilon\to 0$, except for coefficients multiplying the six-dimensional double pentagon master integral defined in reference~\cite{Henn:2021cyv}, whose coefficients are generally of $\mathcal{O}(1/\epsilon)$. These can however be shown to be harmless, since this integral is $\mathcal{O}(\epsilon^2)$, hence its contributions to the finite part of an amplitude vanish, even when multiplied by a pole in the dimensional regulator.

So let us turn to the basis integrals employed in ref. \cite{Henn:2021cyv}
and take their four-dimensional limit.
This is straightforward thanks to the good choice of local integrands \cite{Arkani-Hamed:2010pyv} made in \cite{Henn:2021cyv}..
In fact, the UT integral $I_4^\text{DP}$, a finite six-dimensional double pentagon integral, starts contributing at weight six only, i.e. at order $\epsilon^2$, so we can discard it \footnote{In this paper, we use the notation $I^\text{DP}_i$, $i = 1, \ldots 5$ for
the $I^\text{DP-a}_i$ defined in \cite{Henn:2021cyv}. Note that in \cite{Henn:2021cyv}, the UT integrals
are defined without the overall factor $\epsilon^4$. To extend these
top sector UT integrals  to a whole family UT basis,
the $\epsilon^4$  factor is not included in the auciliary file
UTBasis\_dp.txt.}. Similarly, 
$I_1^\text{DP}$, a parity-odd combination of dual-conformal-invariant integrals, is known to be evanescent \cite{Dixon:2011nj}. The integral $I_3^\text{DP}$
is evanescent due to a Gram determinant that vanishes in four dimensions, and due to its good infrared behavior. So these integrals can also be dropped.

It turns out that the remaining integrals are just the parity-even and parity-odd parts of well-known integrals! Indeed, we have
\begin{align}
I_2^{DP} &= - 2  \tilde \Omega_\text{odd} +O (\epsilon)\,, \label{eq:DP1}\\
I_5^{DP}&= 2  \Omega_\text{even}+O (\epsilon)\,,
\label{eq:DP2}
\end{align}
where $\Omega$ and $\tilde \Omega$ are the finite, dual conformal double pentagon integrals defined in \cite{Arkani-Hamed:2010pyv}, and computed in \cite{Dixon:2011nj}. So the double pentagon top sector UT integrals are either evanescent or known from the studies of $\mathcal N=4$ sYM.

Furthermore, we find that $\tilde \Omega_\text{odd}$ and $\Omega_\text{even}$ can also be represented as sub-sector integrals (up to higher weight terms). We discovered these relations from the symbol results (see discussion below), and then verified them numerically. For $\tilde \Omega_{odd}$, there is a simple relation 
\begin{align}
    \tilde{\Omega}_{\text{odd}}=& 
    { 
\tiny
\dfrac{\Delta_6  G\left(
\begin{array}{ccccc}
l_2-p_5-p_6& p_5& p_6& p_2& p_3\\
l_1+p_5+p_6& p_5& p_6& p_2& p_3
\end{array}\right)}{ 4 \epsilon G\left( p_{5},p_{6},p_{2},p_{3}\right) }
\normalsize
\vcenter{\includegraphics[width=0.22\linewidth]{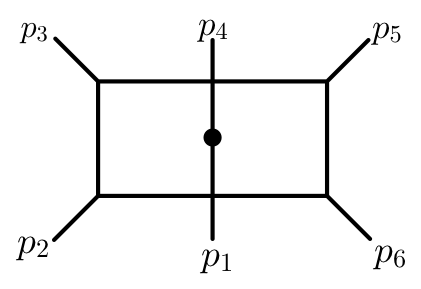} } 
}
 \\
 &\hspace{-1cm} - 
\tiny
 \dfrac{\Delta_6 G\left(
\begin{array}{ccccc}
l_2-p_6& p_5& p_4& p_2& p_1\\
l_1+p_6& p_5& p_4& p_2& p_1
\end{array}\right)}{4 \epsilon G\left( p_{5},p_{4},p_{2},p_{1}\right)} 
\normalsize
{\raisebox{-4ex}{\includegraphics[width=0.22\linewidth]{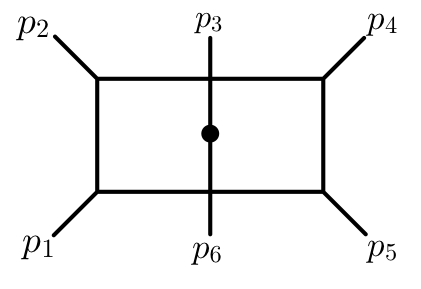}}}
+O (\epsilon) \, , \notag
\end{align}
where the dot indicates a power of $a_{13}=2$.
The weight-four reduction relation for $\Omega_{\text{even}}$ is provided in an auxiliary file. 

\section{Analytic results for the master integrals}

\begin{figure}[t]
    \centering
    \begin{subfigure}[b]{0.25\linewidth}
         \centering
         \includegraphics[width=\linewidth]{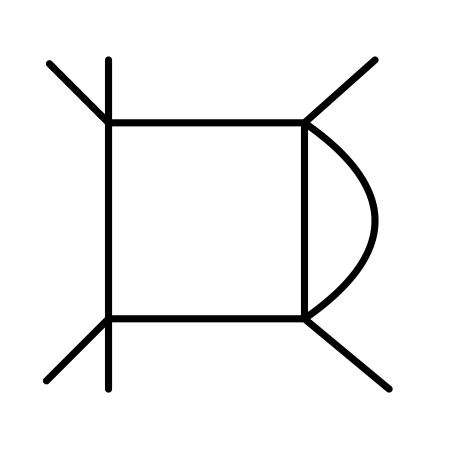}
         \caption{}
         \label{fig:2mdb}
     \end{subfigure}
     \hfill
     \begin{subfigure}[b]{0.25\linewidth}
         \centering
         \includegraphics[width=\linewidth]{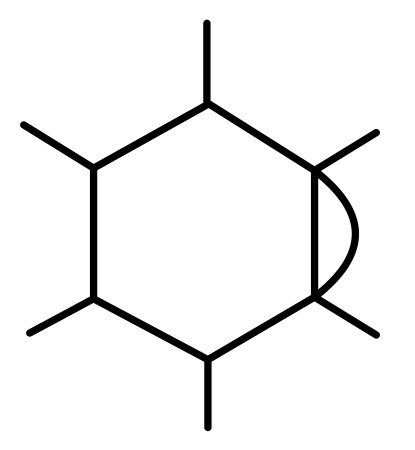}
         \caption{}
         \label{fig:hbb}
     \end{subfigure}
     \hfill
     \begin{subfigure}[b]{0.27\linewidth}
         \centering
         \includegraphics[width=\linewidth]{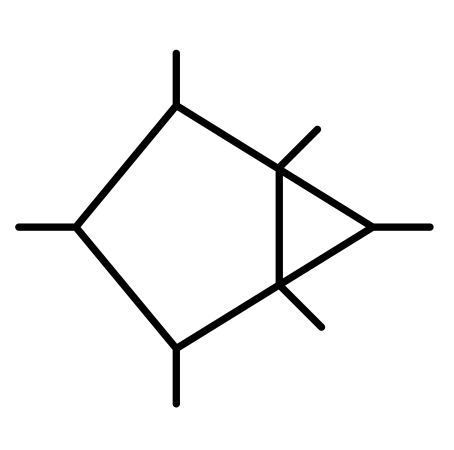}
         \caption{}
         \label{fig:pt}
     \end{subfigure}
     \hfill
     \caption{Integral sectors where genuine six-particle letters 
     first appear.}
     \label{fig:newLetters}
\end{figure}

The canonical differential equations (\ref{eq:CDE}), (\ref{Atildeexplicit}) determine the complete analytic structure of the integrals. Let us now discuss their structure, limiting ourselves to the part up to weight four.

We showed that the top sector of the DP family is not needed for results up to and including weight four in scattering amplitudes. So let us focus on the differential equations for HB, and for all subsectors of DP, as well as any integrals related to them by dihedral symmetry. We find that a total of 245 distinct linearly independent letters appear in them.  

However, as we show below, when solving these differential equations, only a subset of 167 letters appears 
up to a transcendental weight four. The majority of these letters come from five-point sub-sector integrals, known from Ref.~\cite{Abreu:2020jxa}, while only 11 of them are genuine six-particle letters. Starting at weight three, four new letters ($\alpha_{275},\ldots,\alpha_{278})$ appear in the off-diagonal block that relates the hexagon-bubble family, cf. Fig.~\ref{fig:newLetters}(b), to the two-mass-hard box with a bubble insertion, cf. Fig.~\ref{fig:newLetters}(a).
 At weight four, one new letter ($\alpha_{138}$) appears in the sector shown in Fig.~\ref{fig:newLetters}(b), while six new letters ($\alpha_{242},\ldots,\alpha_{247}$) appear in the sector Fig.~\ref{fig:newLetters}(c). 

We use the symbol \cite{Goncharov:2010jf} to describe the full six-particle function space up to transcendental weight four. 
It is straightforward to extract the symbol from the differential equation (\ref{eq:CDE}) at each transcendental weight.
Taking into account dihedral symmetry and modding out linear dependencies, we find the full set of symbols relevant up to weight four. We also include symbols that appear from products of one-loop integrals \cite{Henn:2022ydo}, as such terms are relevant for studies of infrared properties and of the exponentiation of amplitudes.
The results are summarized in Table~\ref{tab:symbol_weights}. The symbols may be useful in future bootstrap approaches. They are given in ancillary files.

Having identified the complete alphabet relevant for four-dimensional Yang-Mills scattering amplitudes, it is interesting to compare it with information that can be obtained from different approaches.
References \cite{Fevola:2023kaw,Fevola:2023fzn} use methods from computational algebraic geometry to analyze the locus of singularities of Feynman integrals. In comparison with our results, we find that this method fails to identify 18 distinct components of the singular locus. Specifically, 15 of these correspond to integrals of the type shown in Fig.~\ref{fig:newLetters}(c), and to its sub-sectors, and the remaining three are leading singularities of the top sector integrals of HB, PB and DB.
Reference \cite{Jiang:2024eaj} is based on the Baikov representation of Feynman integrals. We find that it produces all letters except the letter $\alpha_{100}$ that appears in the family in Fig.~\ref{fig:families}(e).

\begin{table}[t]
\centering
\begin{tabular}{|m{5.7cm}|m{.5cm}|m{.5cm}|m{.5cm}|m{.5cm}|}
\hline
Weight  & 1 & 2 & 3 & 4 \\ \hline
\# Symbols  & 9 & 62 & 319 & 945 \\\hline
\# One-loop squared symbols  & 9 & 59 & 221 & 428 \\ \hline
\# Two-loop five-point symbols   & 9 & 59 & 263 & 594 \\ \hline
\# Genuinely two-loop six-point symbols  & 0 & 0 & 3 & 45 \\ \hline
\end{tabular}
\caption{Independent symbol counting. 
The first row corresponds to the total number of independent symbols at each transcendental weight, including products of symbols of one-loop integrals (the number of which is given in the second row). The last two rows correspond to two-loop five-point one-mass symbols, and to genuine two-loop six-point symbols, respectively. 
}
\label{tab:symbol_weights}
\end{table}



Let us now give more details on the analytic solutions of the differential equations (\ref{eq:CDE}).
We write them as a series expansion around $\epsilon=0$,
\begin{equation}
    \vec{I}(\epsilon, \vec{x}) = \sum_{k=0}^{\infty} \epsilon^{k} \vec{I}^{(k)}(\vec{x}),
\end{equation}
where $\vec{I}^{(k)}(\vec{x})$ denotes functions of transcendental weight $k$ given as $k$-fold iterated integrals
\begin{equation}
    \vec{I}^{(k)}(\vec{x}) = \vec{I}^{(k)}(\vec{x}_0)+\int_{\gamma}  \text{d}A(\vec{x}')  \vec{I}^{(k-1)}(\vec{x'}),
\end{equation}
where $\vec{I}^{(k)}(\vec{x}_0)$ is the boundary value at the reference point $\vec{x}_0$. The boundary values are analytically determined at the reference point $\vec{x}_0=\{-1, 1,0,0,1,1,0,0\}$ in momentum twistor space that corresponds to all Mandelstam variables of eq. (\ref{scalarPoincareinvariants}) being equal to $-1$. We do so by imposing the regularity of our basis integrals throughout the Euclidean region~\cite{Henn:2013nsa,Chicherin:2018mue,Henn:2020lye}. 
 At transcendental weights one and two, the solutions are expressed in terms of a basis of classical polylogarithms. We provide them in the ancillary files \verb|Weight2Solutions_<fam>.m|. At transcendental weights three and four, we express the solutions in terms of one-fold integrals over the weight-two functions~\cite{Caron-Huot:2014lda,Gehrmann:2018yef}
\begin{align}
    \vec{I}^{(3)}(\vec{x})&=\vec{I}^{(3)}(\vec{x}_0) + \int_{0}^{1} \text{d} t \dfrac{\text{d} A}{\text{d} t} \vec{f}^{(2)}(t), \notag\\
    \vec{I}^{(4)}(\vec{x})&=\vec{I}^{(4)}(\vec{x}_0)  \\
    +& \int_{0}^{1} \text{d} t \left(\dfrac{\text{d} A}{\text{d} t} \vec{f}^{(3)}(\vec{x}_0) + \left(A(1) - A(t) \right)\dfrac{\text{d} A}{\text{d} t}\vec{f}^{(2)}(t)\right), \notag
\end{align}
where $t$ parametrizes a straight line between the boundary point $\vec{x}_0$ and a point $\vec{x}$. This representation is used for numerical verification of the master integrals at four different points in the momentum twistor parameterization, within the bulk of the Euclidean region,
\begin{equation}
\scriptsize{
\begin{split}
    \vec{x}_1&=
    \left\{-\frac{299}{300}, \frac{221}{200}, \frac{1}{300}, \frac{1}{200}, \frac{53}{50},  \frac{21}{20}, \frac{1}{100}, \frac{1}{700} \right\},  \\
    \vec{x}_2&=\left\{ -\frac{31}{30}, \frac{161}{150}, \frac{43}{875}, -\frac{17}{630}, \frac{10153}{9450}, \frac{243}{250}, -\frac{22}{75},\frac{1}{30}\right\},  \\
    \vec{x}_3&=\left\{ -\dfrac{17}{10}, \dfrac{51}{50}, \dfrac{109}{3670}, -\dfrac{197}{5505}, \dfrac{3799}{3670}, \dfrac{774}{815}, -\frac{3}{20}, \frac{1}{25} \right\}, \\
    \vec{x}_4&=\left\{ -1, \dfrac{9021}{8950}, \dfrac{7}{1250}, -\dfrac{17}{250}, \dfrac{2003}{1790}, \dfrac{466}{475}, -\frac{637}{3580}, \frac{5}{179} \right\}.
    \end{split}
    }
    \normalsize
\end{equation}
The $40$-digit results we computed from one-fold integration agree with those obtained using \textsc{AMFlow}~\cite{Liu:2022chg,Liu:2017jxz}.

For the DP family, all sub-sector integrals overlap with other families, while the two top-sector integrals contributing at weight four are known from the studies of $\mathcal{N}=4$ sYM, see eq.~\eqref{eq:DP1} and eq.~\eqref{eq:DP2}. Therefore, we use integral representations of $\Omega_{\text{even}}$ and $\tilde{\Omega}_{\text{odd}}$ from~\cite{Dixon:2011nj} for numerical validation of the two integrals at the point $\vec{x}_1$. We find perfect agreement with numerical results obtained using \textsc{AMFlow} \footnote{Inspired by~\cite{Bargiela:2024rul}, an $11$-propagator integral representation for two-loop six-point integrals was used to speed up the AMFlow numeric computation.}

The ancillary files contain the definition of all integral families, together with the list of master integrals, and the associated canonical differential equation matrices (together with a definition of the alphabet of integration kernels), as well as the boundary values at a reference point, expanded up to transcendental weight four. This provides all the information needed to derive analytic solutions up to weight four. Furthermore, we provide a proof-of-concept numerical implementation for evaluating the master integrals in part of the Euclidean region. Additionally, for the reader's convenience, we provide a list of all independent symbols that occur in these solutions.
All the ancillary files can be downloaded from the website
\begin{center}
\url{https://bitbucket.org/yzhphy/2l6p_complete_function_space/}.
\end{center}

\section{Conclusion}
In this paper, we analytically calculated all two-loop six-point planar massless integrals and determined the corresponding function space, up to weight four. 
This is a breakthrough in the field of analytic Feynman integral computation provides the relevant information 
for the calculation of planar amplitudes of $2\to4$ massless QCD processes. 
These results allow the community to initiate research on next-to-next-to-leading order (NNLO) precision physics with four massless final states.

Building upon recent previous work~\cite{Henn:2021cyv,Henn:2024ngj,Matijasic:2024too},
we provided the complete UT basis for all planar two-loop six-point integral families.
We determined the analytic boundary values and expressed the UT integrals as Chen iterated integrals. 
Furthermore, we showed that the double pentagon top sector's UT integrals, up to weight four, can be directly obtained from known functions~\cite{Dixon:2011nj} in the maximally-supersymmetric theory.  
The complete alphabet and the symbol space for two-loop six-point massless integrals are provided, which can be
 used for future bootstrap applications.  
 The fully analytic information we provide can also be used to study interesting physical limits of six-particle scattering processes, such as multi-Regge \cite{Fadin:1993wh} or collinear limits (see e.g. \cite{Dhani:2023uxu}), or the double parton scattering limit of ref. \cite{Dixon:2016epj}, for example.

Another possibility opened up by our work, together
with recent progress on three-loop five-point Feynman integrals \cite{Liu:2024ont},
is to envisage NNNLO cross-section computations for three final states.
This requires extending our results for the master integrals beyond weight four, including the identification of possibly new integration kernels for the double pentagon.
Such higher-weight terms are also required to study the infrared structure of scattering amplitudes at higher loop orders.

{\it Note added:} While this manuscript was in the final
stage of preparation, the preprint~\cite{abreu:2024fei} appeared, which overlaps with some of the results presented here.

\section{Acknowledgments}

It is a pleasure to thank Piotr Bargiela, David Kosower, Sebastian Pögel, Mao Zeng and Simone Zoia for enlightening discussions.
Funded by the European Union (ERC, UNIVERSE PLUS, 101118787 and ERC Starting Grant FFHiggsTop, 101040760). Views and opinions expressed are however those of the authors only and do not necessarily reflect those of the European Union or the European Research Council Executive Agency.  Neither the European Union nor the granting authority can be held responsible for them. Yingxuan Xu is funded by the Deutsche Forschungsgemeinschaft (DFG, German Research Foundation) – Projektnummer 417533893/GRK2575 “Rethinking Quantum Field Theory”. Yang Zhang is supported from the
NSF of China through Grant No. 12075234, 12247103,
and 12047502 and thanks the Galileo Galilei Institute
for Theoretical Physics for the hospitality and the INFN
for partial support during the completion of this work.
This research was supported by the Munich Institute
for Astro-, Particle and BioPhysics (MIAPbP) which is
funded by the Deutsche Forschungsgemeinschaft (DFG,
German Research Foundation) under Germany´s Excellence Strategy – EXC-2094 – 390783311.

\bibliographystyle{h-physrev}
\bibliography{sixpoint.bib}

\begin{thebibliography}{10}

\bibitem{Goncharov:2010jf}
A.~B. Goncharov, M.~Spradlin, C.~Vergu, and A.~Volovich,
\newblock Phys. Rev. Lett. {\bf 105}, 151605 (2010), 1006.5703.

\bibitem{zoia2022modern}
S.~Zoia,
\newblock {\em Modern Analytic Methods for Computing Scattering Amplitudes:
  With Application to Two-Loop Five-Particle Processes} (Springer Nature,
  2022).

\bibitem{Gehrmann:2015bfy}
T.~Gehrmann, J.~M. Henn, and N.~A. Lo~Presti,
\newblock Phys. Rev. Lett. {\bf 116}, 062001 (2016), 1511.05409,
\newblock [Erratum: Phys.Rev.Lett. 116, 189903 (2016)].

\bibitem{Gehrmann:2018yef}
T.~Gehrmann, J.~M. Henn, and N.~A. Lo~Presti,
\newblock JHEP {\bf 10}, 103 (2018), 1807.09812.

\bibitem{Abreu:2018aqd}
S.~Abreu, L.~J. Dixon, E.~Herrmann, B.~Page, and M.~Zeng,
\newblock Phys. Rev. Lett. {\bf 122}, 121603 (2019), 1812.08941.

\bibitem{Chicherin:2018old}
D.~Chicherin {\em et~al.},
\newblock Phys. Rev. Lett. {\bf 123}, 041603 (2019), 1812.11160.

\bibitem{Chicherin:2020oor}
D.~Chicherin and V.~Sotnikov,
\newblock JHEP {\bf 20}, 167 (2020), 2009.07803.

\bibitem{Abreu:2021oya}
S.~Abreu, F.~Febres~Cordero, H.~Ita, B.~Page, and V.~Sotnikov,
\newblock JHEP {\bf 07}, 095 (2021), 2102.13609.

\bibitem{Chawdhry:2021mkw}
H.~A. Chawdhry, M.~Czakon, A.~Mitov, and R.~Poncelet,
\newblock JHEP {\bf 07}, 164 (2021), 2103.04319.

\bibitem{Agarwal:2021vdh}
B.~Agarwal, F.~Buccioni, A.~von Manteuffel, and L.~Tancredi,
\newblock Phys. Rev. Lett. {\bf 127}, 262001 (2021), 2105.04585.

\bibitem{Czakon:2021mjy}
M.~Czakon, A.~Mitov, and R.~Poncelet,
\newblock Phys. Rev. Lett. {\bf 127}, 152001 (2021), 2106.05331,
\newblock [Erratum: Phys.Rev.Lett. 129, 119901 (2022)].

\bibitem{Canko:2020ylt}
D.~D. Canko, C.~G. Papadopoulos, and N.~Syrrakos,
\newblock JHEP {\bf 01}, 199 (2021), 2009.13917.

\bibitem{Chicherin:2021dyp}
D.~Chicherin, V.~Sotnikov, and S.~Zoia,
\newblock JHEP {\bf 01}, 096 (2022), 2110.10111.

\bibitem{Kardos:2022tpo}
A.~Kardos, C.~G. Papadopoulos, A.~V. Smirnov, N.~Syrrakos, and C.~Wever,
\newblock JHEP {\bf 05}, 033 (2022), 2201.07509.

\bibitem{Abreu:2023rco}
S.~Abreu {\em et~al.},
\newblock Phys. Rev. Lett. {\bf 132}, 141601 (2024), 2306.15431.

\bibitem{Badger:2021nhg}
S.~Badger, H.~B. Hartanto, and S.~Zoia,
\newblock Phys. Rev. Lett. {\bf 127}, 012001 (2021), 2102.02516.

\bibitem{Dlapa:2023cvx}
C.~Dlapa, M.~Helmer, G.~Papathanasiou, and F.~Tellander,
\newblock JHEP {\bf 10}, 161 (2023), 2304.02629.

\bibitem{Fevola:2023kaw}
C.~Fevola, S.~Mizera, and S.~Telen,
\newblock Phys. Rev. Lett. {\bf 132}, 101601 (2024), 2311.14669.

\bibitem{Fevola:2023fzn}
C.~Fevola, S.~Mizera, and S.~Telen,
\newblock Comput. Phys. Commun. {\bf 303}, 109278 (2024), 2311.16219.

\bibitem{Golden:2013xva}
J.~Golden, A.~B. Goncharov, M.~Spradlin, C.~Vergu, and A.~Volovich,
\newblock JHEP {\bf 01}, 091 (2014), 1305.1617.

\bibitem{Chicherin:2020umh}
D.~Chicherin, J.~M. Henn, and G.~Papathanasiou,
\newblock Phys. Rev. Lett. {\bf 126}, 091603 (2021), 2012.12285.

\bibitem{Henke:2021ity}
N.~Henke and G.~Papathanasiou,
\newblock JHEP {\bf 10}, 007 (2021), 2106.01392.

\bibitem{Caron-Huot:2016owq}
S.~Caron-Huot, L.~J. Dixon, A.~McLeod, and M.~von Hippel,
\newblock Phys. Rev. Lett. {\bf 117}, 241601 (2016), 1609.00669.

\bibitem{Caron-Huot:2023ikn}
S.~Caron-Huot, M.~Giroux, H.~S. Hannesdottir, and S.~Mizera,
\newblock JHEP {\bf 04}, 060 (2024), 2310.12199.

\bibitem{Henn:2024qjq}
J.~Henn {\em et~al.},
\newblock (2024), 2406.14604.

\bibitem{Dixon:2021tdw}
L.~J. Dixon, O.~Gurdogan, A.~J. McLeod, and M.~Wilhelm,
\newblock Phys. Rev. Lett. {\bf 128}, 111602 (2022), 2112.06243.

\bibitem{TKACHOV198165}
F.~Tkachov,
\newblock Physics Letters B {\bf 100}, 65 (1981).

\bibitem{Chetyrkin:1981qh}
K.~G. Chetyrkin and F.~V. Tkachov,
\newblock Nucl. Phys. B {\bf 192}, 159 (1981).

\bibitem{KOTIKOV1991123}
A.~Kotikov,
\newblock Physics Letters B {\bf 267}, 123 (1991).

\bibitem{Bern:1993kr}
Z.~Bern, L.~J. Dixon, and D.~A. Kosower,
\newblock Nucl. Phys. B {\bf 412}, 751 (1994), hep-ph/9306240.

\bibitem{Gehrmann:1999as}
T.~Gehrmann and E.~Remiddi,
\newblock Nucl. Phys. B {\bf 580}, 485 (2000), hep-ph/9912329.

\bibitem{Henn:2013pwa}
J.~M. Henn,
\newblock Phys. Rev. Lett. {\bf 110}, 251601 (2013), 1304.1806.

\bibitem{Arkani-Hamed:2010pyv}
N.~Arkani-Hamed, J.~L. Bourjaily, F.~Cachazo, and J.~Trnka,
\newblock JHEP {\bf 06}, 125 (2012), 1012.6032.

\bibitem{Hodges:2009hk}
A.~Hodges,
\newblock JHEP {\bf 05}, 135 (2013), 0905.1473.

\bibitem{Henn:2024ngj}
J.~M. Henn {\em et~al.},
\newblock JHEP {\bf 08}, 027 (2024), 2403.19742.

\bibitem{Matijasic:2024too}
A.~Matija\v{s}i\'c,
\newblock {\em {Singularity structure of Feynman integrals with applications to
  six-particle scattering processes}},
\newblock PhD thesis, Munich U., 2024.

\bibitem{Badger:2013gxa}
S.~Badger, H.~Frellesvig, and Y.~Zhang,
\newblock JHEP {\bf 12}, 045 (2013), 1310.1051.

\bibitem{Henn:2021cyv}
J.~Henn, T.~Peraro, Y.~Xu, and Y.~Zhang,
\newblock JHEP {\bf 03}, 056 (2022), 2112.10605.

\bibitem{Dlapa:2021qsl}
C.~Dlapa, X.~Li, and Y.~Zhang,
\newblock JHEP {\bf 07}, 227 (2021), 2103.04638.

\bibitem{vonManteuffel:2014ixa}
A.~von Manteuffel and R.~M. Schabinger,
\newblock Phys. Lett. B {\bf 744}, 101 (2015), 1406.4513.

\bibitem{Peraro:2016wsq}
T.~Peraro,
\newblock JHEP {\bf 12}, 030 (2016), 1608.01902.

\bibitem{Abreu:2020jxa}
S.~Abreu {\em et~al.},
\newblock JHEP {\bf 11}, 117 (2020), 2005.04195.

\bibitem{Heller:2019gkq}
M.~Heller, A.~von Manteuffel, and R.~M. Schabinger,
\newblock Phys. Rev. D {\bf 102}, 016025 (2020), 1907.00491.

\bibitem{Matijasic:2024gkz}
A.~Matija\v{s}i\'{c} and J.~Miczajka,
\newblock {\em {Effortless: Efficient generation of odd letters with multiple
  roots as leading singularities}} (In preparation, 2025).

\bibitem{Hannesdottir:2021kpd}
H.~S. Hannesdottir, A.~J. McLeod, M.~D. Schwartz, and C.~Vergu,
\newblock Phys. Rev. D {\bf 105}, L061701 (2022), 2109.09744.

\bibitem{Gehrmann:2000xj}
T.~Gehrmann and E.~Remiddi,
\newblock Nucl. Phys. B Proc. Suppl. {\bf 89}, 251 (2000), hep-ph/0005232.

\bibitem{Note1}
In this paper, we use the notation $I^\protect \text {DP}_i$, $i = 1, \protect
  \ldots 5$ for the $I^\protect \text {DP-a}_i$ defined in \cite
  {Henn:2021cyv}. Note that in \cite {Henn:2021cyv}, the UT integrals are
  defined without the overall factor $\epsilon ^4$. To extend these top sector
  UT integrals to a whole family UT basis, the $\epsilon ^4$ factor is not
  included in the auciliary file UTBasis\protect \_dp.txt.

\bibitem{Dixon:2011nj}
L.~J. Dixon, J.~M. Drummond, and J.~M. Henn,
\newblock JHEP {\bf 01}, 024 (2012), 1111.1704.

\bibitem{Henn:2022ydo}
J.~M. Henn, A.~Matija\v{s}i\'c, and J.~Miczajka,
\newblock JHEP {\bf 01}, 096 (2023), 2210.13505.

\bibitem{Jiang:2024eaj}
X.~Jiang, J.~Liu, X.~Xu, and L.~L. Yang,
\newblock (2024), 2401.07632.

\bibitem{Henn:2013nsa}
J.~M. Henn, A.~V. Smirnov, and V.~A. Smirnov,
\newblock JHEP {\bf 03}, 088 (2014), 1312.2588.

\bibitem{Chicherin:2018mue}
D.~Chicherin {\em et~al.},
\newblock JHEP {\bf 03}, 042 (2019), 1809.06240.

\bibitem{Henn:2020lye}
J.~Henn, B.~Mistlberger, V.~A. Smirnov, and P.~Wasser,
\newblock JHEP {\bf 04}, 167 (2020), 2002.09492.

\bibitem{Caron-Huot:2014lda}
S.~Caron-Huot and J.~M. Henn,
\newblock JHEP {\bf 06}, 114 (2014), 1404.2922.

\bibitem{Liu:2022chg}
X.~Liu and Y.-Q. Ma,
\newblock Comput. Phys. Commun. {\bf 283}, 108565 (2023), 2201.11669.

\bibitem{Liu:2017jxz}
X.~Liu, Y.-Q. Ma, and C.-Y. Wang,
\newblock Phys. Lett. B {\bf 779}, 353 (2018), 1711.09572.

\bibitem{Note2}
Inspired by~\cite {Bargiela:2024rul}, an $11$-propagator integral
  representation for two-loop six-point integrals was used to speed up the
  AMFlow numeric computation.

\bibitem{Fadin:1993wh}
V.~S. Fadin and L.~N. Lipatov,
\newblock Nucl. Phys. B {\bf 406}, 259 (1993).

\bibitem{Dhani:2023uxu}
P.~K. Dhani, G.~Rodrigo, and G.~F.~R. Sborlini,
\newblock JHEP {\bf 12}, 188 (2023), 2310.05803.

\bibitem{Dixon:2016epj}
L.~J. Dixon and I.~Esterlis,
\newblock JHEP {\bf 07}, 116 (2016), 1602.02107,
\newblock [Erratum: JHEP 08, 131 (2016)].

\bibitem{Liu:2024ont}
Y.~Liu {\em et~al.},
\newblock (2024), 2411.18697.

\bibitem{abreu:2024fei}
S.~Abreu, P.~F. Monni, B.~Page, and J.~Usovitsch,
\newblock (2024), 2412.19884.

\bibitem{Bargiela:2024rul}
P.~Bargiela and T.-Z. Yang,
\newblock Phys. Rev. D {\bf 110}, 096019 (2024), 2408.06325.

\end{thebibliography}

\end{document}